\documentclass[twocolumn,aps,prl,amsmath,showpacs]{revtex4}
\usepackage{graphics}
\begin{document}
\newcommand{\x}{{\mathbf X}}
\newcommand{\W}{{\mathbf W}}
\newcommand{\ksi}{{\boldsymbol \xi}^{\mu}}
\newcommand{\bsxi}{\boldsymbol \xi}
\newcommand{\qh}{\hat{q}}
\newcommand{\Qh}{\hat{Q}}
\newcommand{\sgn}{\text{sgn}}
\newcommand{\eps}{\varepsilon}
\newcommand{\al}{\alpha}
\newcommand{\lan}{\langle\langle}
\newcommand{\ran}{\rangle\rangle}
\newcommand{\btau}{{\boldsymbol \tau}}

\def\lim{\mathop{\rm lim}}
\def\extr{\mathop{\rm extr}}
\def\Tr{\mathop{\rm Tr}}

\title
{Cryptography based on neural networks - analytical results}
\author
{Michal Rosen--Zvi$^1$, Ido Kanter$^1$ and Wolfgang Kinzel$^2$ }
\noindent
\affiliation{$^1$ Minerva Center and Department of Physics,
Bar-Ilan University,
   Ramat-Gan, 52900 Israel\\
 $^2$ Institut f\"ur Theoretische Physik,  Universit\"at W\"urzburg, \\
       Am Hubland 97074 W\"urzburg, Germany}

\begin{abstract}
Mutual learning process between two parity feed-forward networks with
discrete and continuous weights is studied analytically, and we
find that the number of steps required to achieve full synchronization
between the two networks in the case of discrete weights
is finite.
The synchronization process is shown to be non-self-averaging and the
analytical solution is based on random auxiliary variables.  The
learning time of an attacker that is trying to imitate one of the
networks is examined analytically and is found to be much longer
than the synchronization time.
Analytical results are found to be in agreement with simulations.
\end{abstract}

\maketitle

The study of neural networks was originally driven by its potential
as a powerful learning and memory machine. Statistical mechanics
methods have been used to analyze the network's ability and explore
its limitations \cite{Herz,EnvB}. In a recent paper
\cite{KanterKinzel}, the bridge between the theory of neural
networks and cryptography was established.  It was shown numerically
that two randomly initialized neural networks with one layer of
hidden units (so called Parity Machines (PMs)) learning from each
other, are able to synchronize.  The two parties have common inputs
and they exchange information about their output.  In the case of
disagreement, the two PMs are trained by the Hebbian learning rule on
their mutual outputs and they  develop a full synchronized state of
their synaptic weights. This synchronization procedure can be used
to construct an ephemeral key exchange protocol for the secure
transmission of secret data.  An attacker, who knows the
architecture of the two parties, the common inputs, and observes the
mutual exchange of information, finds it difficult to imitate the
moves of the parties and to reveal the
common parameters after synchronization.
All parties have secret informations which are not known
neither to other members nor to possible attackers: Their initial
weights and the current state of their hidden units,
which we are noted as internal representations (IRs)

During the last decade, the analysis of learning from examples
performed by feed-forward multi-layered networks was exhaustively examined
using  statistical mechanics methods \cite{Herz,EnvB}.
An interesting network belonging to this class
is the tree PM which is characterized by a superior
capacity,  as was found by replica calculations \cite{pm}.
The study of the generalization ability of such networks was based on
a set of training examples generated by a static teacher network.
Here we discuss a case where two or several
multilayer networks are trained by their mutual outputs. This
scenario has been solved only for perceptrons and only for continuous ones
\cite{MetzlerKinzelKanter}. Here we present an analytic
solution for PMs with continuous as well as with discrete weights.

In our cryptosystem, each party in the secure channel is represented
by a feed-forward network consisting of $KN$ random input elements
$x_{ji}=\pm 1,~j=1,...,N$, $K$ binary hidden units $\tau_i = \pm 1,
i=1, ..., K$ and one binary output unit $\sigma=\Pi_i \tau_i$.  For
the simplicity of the calculations presented below we concentrate only
on the case of a tree PM with $3$ binary hidden units feeding a binary
output $\sigma=\tau_1 \tau_2 \tau_3$.  The hidden units are determined
via Boolean functions $\tau_i=\sgn(\sum_j W_{ji} x_{ji})$ through
three disjointed sets of inputs $\x_i=x_{1i},...,x_{Ni}$.  The weights
are either discrete or continuous, and the analytical results are
derived for $N \gg 1$.

In this Letter we present: (a) An analytical solution of the mutual
learning of two PMs whose weight vectors are updated according to the
mismatch between their mutual information - their outputs.
Synchronization is achieved in the case of discrete weights,
$W_{ji}=0, \pm 1, ..., \pm L $, as well as
for continuous weights confined to a sphere, $\sum_{j=1}^N W_{ji}^2=N$.
(b) Analysis of online adaptation of discrete weights, in which each
change of a component is not infinitesimally small, demands different
methods than the standard ones \cite{PercBiehl}, and this is at the
center of the discussion below. Surprisingly, synchronization is
achieved for the discrete weights at a finite number of steps. (c)
Dynamical evolution of the discrete networks cannot be characterized by the
time evolution of the standard order parameters,
since the overlaps between the weight-vectors are not self
averaging \cite{Reents} even for large networks.  The analytical
solution is based on calculation of the evolution of the
{\it distribution} of the order parameters as a function of the initial set
of the weights.
(d) The analysis is extended to include
a possible attacker.

For simplicity of presentation, we first describe the
analytical methods developed for the discrete case where detailed
results are presented for particular examined cases.
At the end of this Letter results for the continuous case are also
briefly summarized.

The definition of the updating procedure between the two parties, $A$
and $B$, that are trying to synchronize their weights, is as follows.
In each time step, output of the two parties is calculated for a common
random input.  Only weights belonging to the one (or three) hidden
units which are equal to their output unit are updated, in each one of
the two parties. The updating is done according to the following
Hebbian learning rule,
\begin{eqnarray}
\label{W}
W_{ji}^{A+}=W_{ji}^{A}+ K(W_{ji}^{A}x_{ji} \sigma^{B})x_{ji} \sigma^{B}
\theta(\sigma^A \tau_i^A) \theta(-\sigma^A\sigma^{B}), \\ \nonumber
W_{ji}^{B+}=W_{ji}^{B}+ K(W_{ji}^{B}x_{ji} \sigma^{A})x_{ji} \sigma^A
\theta(\sigma^B \tau_i^B)\theta(-\sigma^A\sigma^{B}),  \nonumber
\end{eqnarray}
where $K(y)=1-\delta_{L,y}$ and $\delta$ represents the Kronecker
function. The purpose of the operator $K(y)$ is to prevent the
increment (decrement) of the strength of the weights on the boundary
value $L$($-L$).

Two important simulation results  are crucial for the analytical description
of the mutual dynamics.
The first observation is that the synchronization time
is finite \cite{KanterKinzel}.
The second
is that different runs (set of random inputs) of the above
dynamics, but with fixed initial conditions for the two parties, result in
different sets of IRs. As a result of these two
observations, we realized that the variance of the overlaps between the
two parties is finite and does not shrink to zero even in the
thermodynamic limit.  This unusual scenario of on-line mutual learning
is taken into consideration in the analytical equations, by the
selection of random IRs following the freedom given by the current analytical
overlaps. We find an iterative
discrete set of equations for the mutual overlaps between the parties,
whose evolution depend on some random but correlated ingredients - the
current IRs, $\{\tau_i^A\},\{\tau_i^{B}\}$
(see Eq. \ref{W}).

In each time step, $\mu$, the mutual state of the two parties is
defined by a $(2L+1)\times( 2L+1)$ matrix,
$\overline{\overline{F^i}}(\mu)$, where $i$ represents the hidden
unit.  The element $f_{qr}^i$ of the matrix stands
for the fraction of components in the $i$th weight-vector which are
equal to $q(r)$ in the first(second) party, where
$q,r=0, \pm 1, ..., \pm L$.  The overlap of the weights belonging to
the $i$th hidden unit in the two parties,
$R_i^{A,B}=\W_i^A \cdot \W_i^{B}/N$, as well as their
 norms, $Q_i=\W_i \cdot \W_i$/N,  are
given by the matrix elements
\begin{equation}
\nonumber
R_i^{A,B}=\sum_{q,r} qr f^i_{qr},~~
Q_i^{A}   =\sum_{q,r} q^{2} f^i_{qr},~~
Q_i^{B}  =\sum_{q,r} r^{2} f^i_{qr}.
\end{equation}
These overlaps and norms fixed the probabilities of deriving the same
IR via the normalized overlap,
$\rho_i^{A,B}=R_i^{A,B}/{\sqrt{Q_i^A Q_i^{B}}}$.
More precisely, the probability of having different results in the
$i$th hidden unit of the two parties is given by the
well known generalization error for the perceptron
$\epsilon_p^i=\cos^{-1}{\rho _i}/{\pi}$ \cite{Herz,EnvB}.

Each of the PM consists of a tree architecture and for random inputs
each of the $8$ IRs appears with equal probability. The joint
probability distribution of the $64$ different pairs of IRs in both
parties is correlated, and can be explicitly expressed using
$\{\epsilon_p^i\}$.

The development of
the elements of the matrix $\overline{\overline{F^i}}(\mu)$ are calculated
directly from Eq. \ref{W}, where one has to average over the inputs
$x_{ij}$. We use auxiliary random variables in order to choose one of
the possible IRs following their probabilities given by
$\{\epsilon_p^i\}$.  In each step we choose two sets of random
numbers which are taken from a flat distribution between $0$ and $1$:
Set I: In the event that the number is smaller than $\epsilon_i$ we
deduce that the two hidden units disagree, otherwise we assume an
agreement.  Set II: All eight IRs are equally probable in the first
party, since the architecture consists of a tree PM. We choose one
among the eight using the second set of auxiliary variables $p_r$, and the
corresponding IR for the second network according to the first set.

To exemplify derivation of the iterative equations for
$\{f^i_{qr}\}$, let us concentrate on the case
where the result of the first random set is that all three hidden units
are in disagreement.
In two possibilities out of the eight IRs all
three hidden units are updated, whereas in the other six
possibilities only one is updated (we then have to choose at
random one among the three).  After taking into account all possible
internal scenarios, and accordingly the updates, one can show that the
iterative equations for $\{f^i_{qr}\}$ away from the boundary, $q,r \ne
\pm L$, are given by
\begin{eqnarray}\nonumber
f_{q,r}^{i+} =\theta(\frac{1}{4}-p_r)(\frac{1}{2}f^i_{q+1,r-1}+
\frac{1}{2}f^i_{q-1,r+1})+\\  \nonumber
\theta(\frac{i+1}{4}-p_r)\theta(p_r-\frac{i}{4})(\frac{1}{2}f^i_{q+1,r-1}+
\frac{1}{2}f^i_{q-1,r+1}). \nonumber
\end{eqnarray}
On the boundary, similar equations can be derived as well as for other
internal scenarios. Taking into account all possible scenarios and the
inversion symmetry of our PMs,
one has to solve iteratively only $4$ classes of equations in a manner
similar to the abovementioned \cite{long_paper}.
Note that the time evolution of the
$f_{qr}$ and the overlaps depends on time dependent random variables.

\begin{figure}[h]
\centering {
\resizebox*{1\columnwidth}{!}{
\rotatebox{270} {
\includegraphics{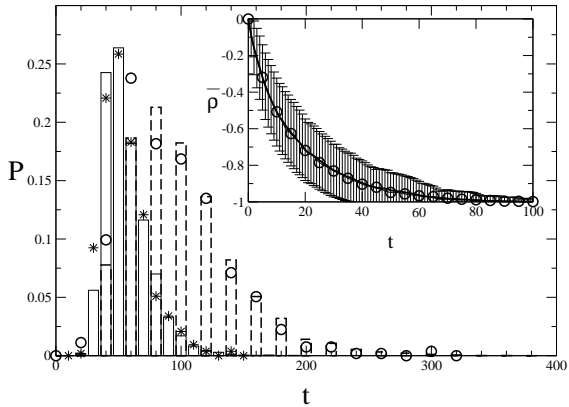}
}}}
\caption { The histogram of the $t_{synch}$ (solid line) and
$t_{learn}$ (dashed line) as was obtained in different runs of the
discrete iterative equations for PMs with $L=1$. Symbols
stand for simulation results, $N=10000$ based on $500$ runs.
Inset: numerical results of $\overline{\rho}$ as a function of the
number of steps. Analytical results (solid line) and simulations
results (circles) include the standard deviation
obtained from $500$ different runs.}
\label{dist1}
\end{figure}

Different runs for updating of the equations result in different
trajectories of the order parameters. In the inset of Fig.
\ref{dist1}, we present the average overlap
$\overline{\rho}=\sum_{i=1}^3\rho_i/3$, and its standard deviation,
obtained from $500$ different runs of  the analytical equations with $L=1$.
Results of the averaged overlap (with the same standard deviation) obtained 
in $500$ runs of simulations with $N=10^4$ are denoted by circles.

An important quantity is the number of steps required to achieve full
synchronization, $t_{synch}$, since it  can be used by the parties
to encrypt/decrypt the information using the known output bit.  In
simulations the synchronization time is well defined - the first step in
which all weight vectors of the parties are in an anti-parallel state.
In contrast, in the analytical solution the average overlap of the
hidden units tends to zero exponentially with the number of steps. In
order to compare analytical results to simulations we need to find a
criterion which determines synchronization.  We chose the criterion
$\overline{\rho} \le -c_l=-1+0.1/(NL)$ to define full
synchronization, since $c_l$ is much greater than the maximal possible
overlap just before synchronization.

The exponential decay of the overlaps with the number of steps and the
claim that synchronization is achieved at a finite number of steps
even for $N \gg 1$ has to be clarified.  Our synchronization process
is mainly characterized by two regimes: The first $t_a$ steps which 
are characterized by different IRs (in some of the steps) for the two
parties. Note that $t_a$ is fluctuating from sample to sample.  The second is 
the asymptotic regime, 
last $t_b$ steps, where the IRs of the parties are always the
same, and the weights are converging to an anti-parallel state
similarly to three perceptrons, $t_b \propto \log(N)$
\cite{long_paper}.  Roughly speaking, the two regimes are
characterized by $\epsilon_i^P > 1/t_a$ and $\epsilon_i^P < 1/t_b$,
respectively.  Our analytical results as well as simulations indicate
that $t_a$ is independent of $N$. Hence as long as $t_a > t_b$, the
$\log(N)$ dependent is invisible.  For $L=3$, for instance, $t_{synch}
\sim 400$, $t_a \sim 300$, and $t_b$ is expected to be equal to $t_a$
only for $N \sim e^{200}$.

In Fig. \ref{dist1}, we present the histogram of the number of steps
required to achieve $t_{synch}$, $P(t_{synch})$, in simulations with
$N=10^4$ and $L=1$ and the initial weights were chosen such that
$\rho^{A,B}_i=0$. This distribution is in a fairly good agreement with the
results obtained by the runs of the iterative equations for
$f_{qr}$.

Let us now examine a possible attack of a third player, an attacker
$o$, that tries to imitate one of the parties (let us say $A$).  We
assume that the attacker uses the same algorithm as one of the
partners.  The attacker updates its own weight-vectors only when an
updating step is taken by the parties. The natural move of an
attacker in such an event is to follow the rule of the parties
\begin{equation}
W_{ji}^{o+}=W_{ji}^{o}+K(W_{ji}^{o}x_{ji}\sigma^{B})x_{ji}\sigma^{B}
\theta(\sigma^A \tau_i^o)\theta(-\sigma^A\sigma^{B}),
\nonumber
\end{equation}
indicating that only weight-vectors belonging to the hidden units
which are in agreement with the output of party $A$ are updated, (
more advanced attacks will be discussed elsewhere \cite{long_paper}).
The evolution of the overlap of an attacker depends on the evolution
of $6$ matrices; three matrices describing the overlaps between the
parties and similarly, three matrices describing the overlaps between
the attacker and the first party.  Note that the dynamics of the
attacker depends on moves of the parties which depend on their
overlaps. Hence, the time evolution of six matrices gives the full
description of the overlaps between the attacker and the first party
and between the parties themselves.  The mutual dynamics of the three
networks, two parties and the attacker, depends on the joint
probability distribution of $8 \times 8 \times 8$ IRs, and upon the
corresponding updates of the six matrices. The full description of the
discrete time evolution of the matrices and the overlaps will be given
elsewhere \cite{long_paper}.

The analytical solution of the dynamics of the attacker indicates that
a full learning is achieved in a finite number of steps, $t_{learn}$, 
where a full learning is defined such that 
$\overline{\rho}^{A,o}>c_l$.  In Table
\ref{tab} $t_{learn}$ and $t_{synch}$ are compared for various $L$.

\begin{table}[h]

\label{tab}
\begin{tabular}{|c|c|c|c|}
\hline
&
\( t_{synch} \)&
\( t_{learn} \)&
\( r \)\\
\hline
\hline
L=1&
\( 61\pm 10 \)&
\( 1.1\cdot 10^{2}\pm 0.2\cdot 10^{2} \)&
\( 1.8\pm 0.6 \)\\
\hline
L=2&
\( 188\pm 26 \)&
\( 1.5\cdot 10^{3}\pm 0.5\cdot 10^{3} \)&
\( 8.0 \pm 2.9 \)\\
\hline
L=3&
\( 376\pm 51 \)&
\( 4.5\cdot 10^{4}\pm 1.3\cdot 10^{4} \)&
\( 120 \pm 51 \)\\
\hline
L=4&
\( 673\pm 95 \)&
\( 6.9\cdot 10^{7}\pm 5.7\cdot 10^{7} \)&
\( 1.04\cdot 10^{5}\pm 1.02 \cdot 10^{5} \)\\
\hline
\end{tabular}
\caption {The average synchronization time, $t_{synch}$, the average
learning time $ t_{learn}$, their standard deviation  and the ratio
$t_{learn}/t_{synch}$ averaged over $2000$ different runs of the
iterative equations with the halting criterion $c_l=1-10^{-5}$.}

\end{table}

For $L=1$ the average learning time is about twice the synchronization
time, and one may reach the wrong conclusion that the
synchronization process always terminates before the learning process.
In Fig. \ref{dist1} we present the histogram of the
synchronization and the learning processes,  and
a fairly good fit between analytical and simulation results is apparent.
The two distributions,
$P(t_{synch}), ~P(t_{learn})$ have a finite overlap,
indicating that in a finite fraction of the runs the learning process
terminates before the achievement of synchronization (which was indeed
observed in a finite fraction of the runs of the simulations). Hence
the construction with $L=1$ is not a good candidate to build a
secure channel.

\begin{figure}[h]
\centering {
\resizebox*{1\columnwidth} {!} {
\rotatebox{270} {
\includegraphics{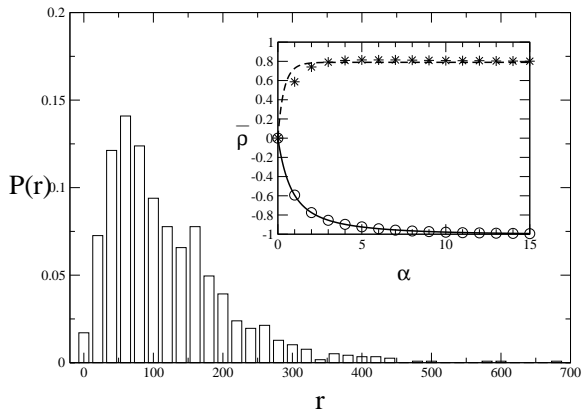}
}}}
\caption { The distribution of $r=t_{learn}/t_{synch}$ for $L=3$
obtained from the analytical solution of about $1200$ runs. The lowest
value obtained for $r$ was $\sim 6$.  Inset: The average overlaps
$\overline{\rho}^{A,B}$ (solid line) and $\overline{\rho}^{A,o}$
(dashed line) as a function of $\alpha$ for PMs with continuous
weights and $\eta=3$ are presented.  Symbols stand for simulation
results with $N=5000$ and error bars are smaller than the symbols.  }
\label{ratio}
\end{figure}

For $L \ge 3$ the ratio $r=t_{learn}/t_{synch}$ averaged over the runs
was found to be $r \gg 1$ (see Table \ref{tab}).  For $L=3$, we did
not observe, in simulations over $10^5$ runs, a case where $t_{learn}$
was faster than $t_{synch}$. In
Fig. \ref{ratio} we present the histogram of the probabilities of the ratio,
$r$, as was found by averaging over different runs
of the analytical equations.  The minimal value of the ratio was 
$r \sim 6$ where the largest ratio was $r \sim 680$.  We found that the
largest synchronization times are smaller than $1000$ whereas the
typical learning time is $4.5\cdot 10^{4}$.

Synchronization in the case of PMs with continuous weights is
achievable only with the following modifications. (a) Normalization
of the weight vectors belonging to each one of the hidden units after
every updating step. The natural normalization we use is the spherical
normalization, $\sum_{j=1}^N W_{ji}^2=N$.
(b) The change in the strength of each weight (before
normalization) is $\eta/N$, where $\eta$ is a constant of order one.
The synchronization time is proportional to the size of the input,
$N$, and therefore the analytical description of the system is given
by a coupled differential equations. Some limited results and brief
description of the method are presented below. More detailed
results will be given elsewhere \cite{long_paper}.

Updating of weights of the first party for the spherical case
is given by
\begin{equation}\label{cont}
 \W_{i}^{A+} = \frac{\W_{i}^{A}+\frac{\eta}{N}\x_{i}
        \theta(-\sigma^A \sigma^{B})
\theta(\sigma^A \tau_i^A)
\sigma^{B}}{|{\W_{i}^{A}+\frac{\eta}{N}\x_{i}
        \theta(-\sigma^A \sigma^{B})
\theta(\sigma^A \tau_i^A) \sigma^{B}}|}   \nonumber
\end{equation}
and similarly the updating rules for the second party and the
attacker.  The analytical calculation can be simplified in the
continuous case by the probability
that there is a mismatch between the two PMs given that
there is a mismatch between two hidden units, $P^{i}_1 \equiv
P(\sigma^A \neq \sigma^{B} | \tau_i^A\neq \tau_i^{B})=\epsilon_p^{j}
\epsilon_p^{k}+(1-\epsilon_p^{j})(1-\epsilon_p^{k})$ and similarly
$P^{i}_2=P(\sigma^A \neq \sigma^{B} |\tau_i^A = \tau_i^{B})=1-P^{i}_1$.
One can map the mutual process onto that of perceptrons, where the updating
of the first party, for instance, is given by
\begin{equation}\label{percept}
\W^{A+}_i = (\W^{A}_i+\frac{\eta}{N}\x_i \tau^{B}\Delta^A_i)
/|\W^{A}_i+\frac{\eta}{N}\x_i \tau^{B}\Delta_i^A| \; \;
\nonumber
\end{equation}
and similarly for the second party, where
$\Delta^A_i =  \theta(-\tau_i^A \tau^{B}_i)\theta(\frac{P_1^{i}}{2}-p_a)
+\theta(\tau_i^A \tau_i^{B})\theta(P_2^{i}-p_b)\theta(\frac{1}{2}-p_c)$
and we use auxiliary variables $p_a$, $p_b$, $p_c$ to specify each run.

The next step consists of the averages over the following two
quantities.  (a) Averaging over the joint probability distributions of
the local fields of the two parties. (b) Average over the auxiliary variables,
 which is unique to the case of mutual learning.
The normalized overlap, $\rho$, between weight vectors belonging to each pair
of hidden units is found to obey the equation,
$d\rho/d\alpha=\eta[C^2+(1-C)^2]((1-\rho)/\sqrt{2 \pi}-\eta C/2)(1+\rho)
-2 \eta(1-\rho^{2})C(1-C)/\sqrt{2\pi} -\eta^2 \rho C(1-C)^2$,
where $C=\cos^{-1}{\rho}/\pi$.
For  $\eta<\eta_c \sim 2.68$ the points $\rho=\pm 1$ are repulsive
fixed points of the above equation, where
for $\eta > \eta_c$ a phase transition occurs to a state of full
synchronization.

The equation of motion of the overlap of an attacker with the first party
after synchronization, i.e., $\rho^{A,B}=-1$,
$\rho^{A,o}=-\rho^{B,o}$, is given by
$d\rho^{A,o}/d\alpha=\eta^2
(1-\cos^{-1}{\rho^{A,o}}/\pi -\rho^{A,o})/2$.
The fixed point of this equation is $\rho^{A,o}= -\rho^{B,o} \sim 0.79$
and is independent of $\eta$, indicating that perfect learning is not
achievable. Analytical results derived from the last two
equations in the case of $\eta=3$ are presented in the
inset of Fig. \ref{ratio} and are in good agreement with simulation
with $N=5000$ and $20$ runs.

I.K. acknowledges the support from the Israel Academy of Science.

\end{document}